# Post Graphene 2D Chemistry: The Emerging Field of Molybdenum Disulfide and Black Phosphorus Functionalization

Andreas Hirsch* and Frank Hauke

[a]   Prof. Dr. A. Hirsch, Department for Chemistry and Pharmacy & Joint Institute of Advanced Materials and Processes (ZMP), Friedrich-Alexander-Universität Erlangen-Nürnberg (FAU), Henkestraße 42, 91054 Erlangen
E-mail: andreas.hirsch@fau.de
[b]   Dr. F. Hauke, Joint Institute of Advanced Materials and Processes (ZMP), Friedrich-Alexander-Universität Erlangen-Nürnberg (FAU), Dr.-Mack-Str. 81, 90762 Fürth

**Abstract**

The current state of the chemical functionalization of three types of single sheet 2D materials, namely, graphene, molybdenum disulfide ($MoS_2$), and black phosphorus (BP) is summarized. Such 2D sheet polymers represent currently an emerging field at the interface of synthetic chemistry, physics, and materials science. Both covalent and non-covalent functionalization of sheet architectures allows for a systematic modification of their properties, i.e. an improvement of solubility and processability, the prevention of re-aggregation or a band gap tuning. Next to successful functionalization concepts also fundamental challenges are addressed. These include the insolubility and polydispersity of most 2D sheet polymers, the development of suitable characterization tools, the identification of effective binding strategies, the chemical activation of the usually rather unreactive basal planes for covalent addend binding, and the regioselectivity of plane addition reactions. Although a number of these questions remain elusive in this review, the first promising concepts to overcome such hurdles have been listed.

## 1. Introduction: 2D-Layered Systems

The materials chemistry of the 20[th] century was dominated by organic polymers. These polymers consisted predominantly of one dimensional (1D) macromolecular chains. The resulting polymeric materials played - and still play - a fundamental role in our modern civilization. Products made from 1D polymers such as teflon, polystyrene, polypropylene, polyacrylates, polyamides, and polyurethanes are indispensable. With the recent advent of two dimensional (2D) materials the 21[th] century seems to witness a new development, which bears great potential for another revolution in materials science. The year 2004 has seen the awakening of the sleeping beauty graphene,[1] a 2D polymer which had been considered to be an exclusively theoretical material[2] due to its predicted thermodynamic instability at any finite temperature.[3] Ever since graphite existed, graphene was there but it was hidden within the stacked arrangement of the individual sheets that form the underlying three dimensional (3D) structure of this carbon allotrope.[4] Only once exfoliated to individual sheets, the remarkable properties of this 2D sheet polymer, which were predicted beforehand[5] and which are fundamentally different from its bulk storage form graphite, could be revealed experimentally. In this context, the observation of graphene's ambipolar field effect,[1a] the quantum Hall effect at room temperature,[6] and its extremely high carrier mobility[7] has to be named. Due to it's unique electrical properties and based on it's 2D-extended surface area, graphene became the first system ever capable of the detection of a single molecule adsorption event.[8]

Once exfoliated, graphene can be supported and stabilized on a substrate which is a prerequisite not only for investigating but also for using its properties for practical applications.[9] Such a substrate storage is very important since it prevents the re-aggregation of the individual layers. This event is unavoidable if, for example, metastable dispersions of graphene in an organic solvent would be allowed to age.[10] Graphene cannot be isolated as a bulk material since at least partially it would always convert back to graphite. The energy associated with the corresponding π-π-interaction of sp$^2$ hybridized carbon surfaces is overwhelming.[11] Fortunately, graphene as an atomically thin 2D material is not condemned to a sole existence. Inorganic text books tell us that there is a whole variety of 3D solid state structures with a blueprint based on stacked 2D layers,[12] i.e. transition metal dichalcogenides (TMDs) such as molybdenum disulfide ($MoS_2$),[13] hexagonal boron nitride,[14] germanium,[15] black phosphorus (BP),[16] and antimony.[17] As for graphene, remarkable properties have been discovered for this novel class of compounds within the last few years. For instance, $MoS_2$ is a frequently used solid state lubricant and catalyst for hydrodesulfurization- and hydrogen evolution reactions.[18] It can be isolated in two polytypes, 2H-$MoS_2$ and 1T-$MoS_2$, with distinct differences in their electronic properties.[19] Thin layered 2H-$MoS_2$ has an electronic bandgap that increases with decreasing layer thickness[20] and is a strong luminophore,[21] whereas 1T-$MoS_2$ is a metallic conductor.[22]

All 2D materials in the focus of this review article (Fig. 1) share one common concept, the individualized single sheets can be prepared *via* exfoliation of the corresponding 3D mother compounds.[10-11, 13, 16b, 23]

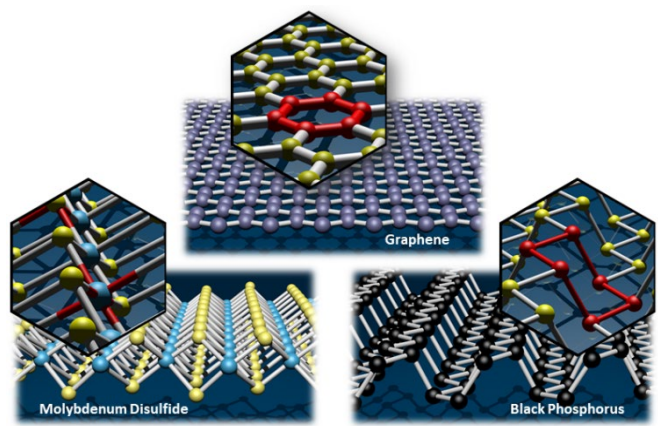

**Figure 1.** 2D-layered materials. a) Graphene – planar six membered sp$^2$ carbon ring highlighted in red. b) Molybdenum disulfide (2H-$MoS_2$) – trigonal pyramidal binding motif of the bridging sulfur atoms and trigonal prismatic coordination of the molybdenum centers are highlighted in red. c) Black phosphorus – six membered phosphorous ring in chair configuration highlighted in red.

For these novel 2D sheet polymers it has been demonstrated that the properties of the individual decoupled sheets differ significantly from those of the constituting 3D solid state bulk materials.[23f] With the achieved progress in top-down compound preparation techniques the door for the investigation of 2D materials with a very broad spectrum of physical and chemical properties has been opened. In the meantime also bottom-up approaches like CVD processes have been developed as alternative production methods, which was successfully exemplified in the case of graphene synthesis.[24]

Being at the starting point of a new research direction it is advisable to define the most suitable terms for the objects under investigation in order to avoid misconceptions. In this review article we will mainly focus on graphite, $MoS_2$, and BP as the corresponding mother 3D solids of the respective 2D sheet polymers materials under consideration. They represent an allotrope of an element (C or P) or a covalently coupled binary compound ($MoS_2$). The terms graphene, single layer molybdenum disulfide (SL-MoS$_2$), and single layer black phosphorus (SL-BP) describe exactly one single layer of these systems. In the case of SL-BP the alternative expression, phosphorene, that appeared recently in literature[23k, 25] is misleading because it implies the presence of double bonds and a lattice of sp$^2$ hybridized P atoms. For the description of bilayer BP/MoS$_2$ and the respective few layer analogues, consisting of 3-10 stacked layers, we will use the term BL-BP/MoS$_2$ and FL-BP/MoS$_2$, respectively.

The common feature of all single layer 2D sheet polymers is their atomically thin structure. They exist only on surfaces or they are stabilized by physisorbed molecules or ionic liquids. Transparency is combined with mechanical flexibility and opto-electronic activity. Furthermore, many representatives such as graphene and SL-MoS$_2$ are very stable under ambient conditions. The combination of a very high charge carrier mobility,[26] good electrical conductivity, and a extraordinary high stability found in graphene is absent in all classical 1D conjugated polymers, i.e. polyacetylenes.[27] In the latter systems a high conductivity can only be accomplished after doping,[28] which on the other hand renders these materials rather unstable. As a consequence, an enormous potential for exciting applications of 2D sheet polymers arises. Beyond this, one can even think about combining the different material types in heterostructure assemblies, an almost unlimited playground for scientists and materials engineers.[29]

The required materials engineering for a controlled device production, however, is associated with a variety of challenges. These include the large scale production, solution and/or melt processing, and the facile tuning of physical and materials properties. This is where chemistry comes into play. The chemical modification and functionalization of 2D sheet polymers not only allows for the modification of their properties but also for their combination with the property portfolio of other compound classes. From a more fundamental scientific perspective it is of course also of great interest to study the inherent chemical reactivity principles as such. The deduction of the underlying reactivity patterns will be of great value for the systematic design for the whole range of 2D sheet polymer-based technological applications.

We will present here the most important concepts of 2D sheet polymer chemistry. Whereas the functionalization of graphene has already been investigated since a few years,[30] the chemical functionalization of other representatives has just started.[31] Here, we will focus mainly on the chemistry of MoS$_2$ and BP. First, we will discuss and evaluate important concepts with respect to their non-covalent and covalent functionalization. Our own group has contributed to these developments in the same chronological order, namely, starting with graphene chemistry[30b, 32] (which in turn is based on our experience with fullerene[33] and carbon nanotube chemistry[34]) over the functionalization of MoS$_2$,[31c] and BP.[23l, 35]

The chemistry of graphene has been reviewed quite extensively.[30] Within this article we will just point out the most important concepts, challenges, and problems. We will rather put the most recent pioneering contributions in MoS$_2$ and BP functionalization at the forefront, serving also as role models for other 2D sheet polymers so far, being less explored.

## 2. General Remarks

Before we start with the description of the case studies, as pointed out above, we will analyze a couple of general problems and challenges associated with 2D materials chemistry (Fig. 2). The first challenge is certainly the structural characterization of the chemically functionalized graphene-, MoS$_2$-, BP-, and 2D sheet polymer materials in general. Reaction products obtained from a conventional chemical conversion of discrete molecules can usually be purified, for example, by chromatography, recrystallization or distillation. Subsequently, they can usually be unambiguously and quickly be characterized using routine analytics. With a mass spectrum, NMR characterization, elemental analysis or even X-ray crystallography at hand the structure of a compound can usually be exactly determined. However, in the case of 2D sheet polymer functionalization these powerful purification and characterization methods cannot be applied. Moreover, it is an intrinsic feature of the 2D sheet structures - they truly represent 2D polymers - that they are polydisperse and exhibit a certain size-, shape-, and polymer weight distribution. Moreover, the corresponding reaction products obtained from covalent addition reactions are also far from being uniform, what the degree of functionalization and the regiochemistry (addition patterns) is concerned.

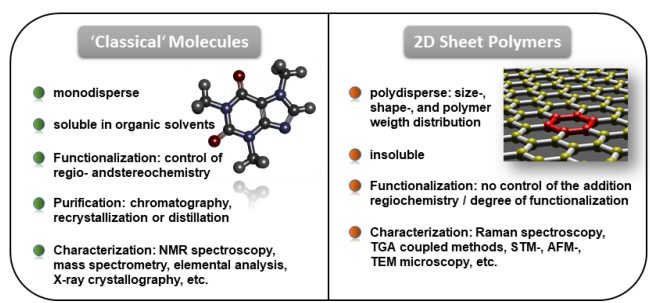

**Figure 2.** Challenges of 2D sheet polymer chemistry and characterization.

The challenge is now to develop and provide analytical tools beyond the standard methods commonly used by synthetic chemists. An important aspect is the statistical relevance of such methods. Due to the potential sample inhomogeneity and possible inhomogeneous regiochemistry of addition reactions (addend clustering in certain areas), single measurements at certain areas of a sample can be strongly misleading. For this purpose we have recently introduced statistical Raman spectroscopy (SRS) and statistical Raman microscopy (SRM) for the 2D materials characterization.[36] This was a major breakthrough in the field since the fast analysis of serval thousands of spatially resolved Raman spectra allowed for a representative characterization of functionalized 2D polymer flakes and a suitable representation of the results, for example, by property-2D-maps or histograms. High resolution transmission electron microscopy (HRTEM) provides the advantage of atomic resolution but is of limited analytical value for chemically functionalized samples, since a) the high energy electron beam can easily lead to a defunctionalization of the sample and b) the statistical information required for a quantitative analysis with respect to the degree of addition is still difficult to extract.[37] The latter is also true for scanning probe microscopy. The combination of thermogravimetry (TGA) coupled to mass spectrometry and gas chromatography turned out to be a rather efficient tool for the quantitative and qualitative (nature of added groups) analysis in the course of the covalent functionalization/analysis of 2D-layered systems.[38] Solid state NMR spectroscopy has been applied successfully only in a very limited number of cases.[39] Optical spectroscopy has been applied to investigate a) 2D sheet polymer based fluorescence quenching by addends bound either covalently[40] or non-covalently[35a, 41] and b) discharging of negatively charged graphene sheets in order to quantify the degree of charging.[42] Recently, we succeeded in the *in situ* Raman monitoring of the covalent functionalization of graphene allowing for the identification of functionalization related new Raman bands.[43] However, one has to clearly point out, that by none of these analytical approaches a precise information about the underlying addition pattern or the exact mode of chemical binding can be obtained. In order to access insights into these fundamental questions a comparison of experimental results with those obtained form quantum mechanical model calculation provides the most powerful source of information.[35a, 43]

Another aspect concerns the topicity of addition reactions (Fig. 3).[44] In contrast to the chemistry of closed fullerenes[33] and carbon nanotubes,[45] which can only be attacked from one side (exohedrally), subsequent covalent addend binding to graphene can take place at both sides of the basal plane when the chemical transformation is carried out in homogeneous dispersion[44a, c] or deposited on a reactive substrate.[44b] Such "both-sided" additions are referred to as *antaratopic* addition sequences. If on the other hand the 2D sheet polymer is deposited on an inert substrate, only a "one-sided" or *supratopic* addition sequence is possible. In the case of graphene, subsequent *antaratopic* binding steps can lead to strain free products like, in the extreme case, graphane with an all chair configuration.[39b] On the other hand if only *supratopic* reaction sequences are possible highly strained architectures will be formed being less stable than the starting material.

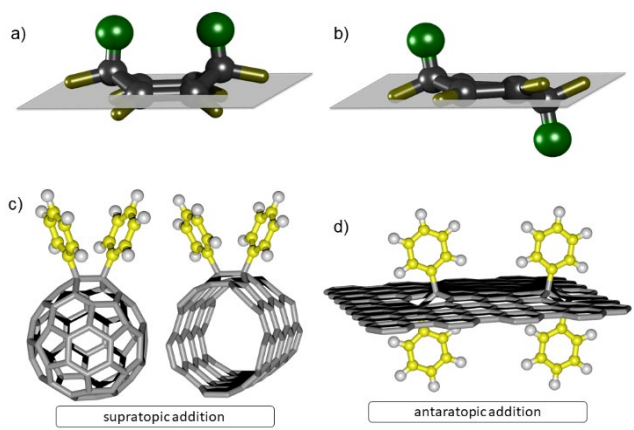

**Figure 3** Topological aspects of carbon allotrope addition reactions. Top: Schematic representation of addend relation in the case of a) a *supratopic* and b) *antaratopic* 1,4-addition reaction. Bottom: c) Exohedral (*supratopic*) functionalization of closed cage carbon allotrope structures – 1,2-addition of phenyl rings. d) 1,2-Addition of phenyl rings in the case of graphene.

Finally, we would like to have a look at the expected chemical reactivity of our model cases graphene, *SL*-MoS$_2$, and *SL*-BP. All these single sheet architectures have a rather high surface energy, which has to be overcome during the exfoliation of their corresponding 3D mother architecture. Conceptually, this surface energy can be lowered if external molecules would be allowed to non-covalently bind to the sheet, *via* π-π- or n-π-interactions. The resulting gain in energy provides the driving force for the non-covalent supramolecular functionalization to these 2D sheet polymers. The concept of the supramolecular stabilization of exfoliated 2D sheets has recently been successfully applied in the non-covalent functionalization of BP (Fig. 4).[35a] Here, the interaction with 7,7,8,8-tetracyano-*p*-quinodimethane (TCNQ) (**1**) and an electron deficient perylene bisimide derivative (**2**) led to a considerable stabilization of the BP flakes against oxygen degradation.[35a] This oxydation passivation can also be achieved by a coverage of the BP surface with ionic liquids.[35b, 46]

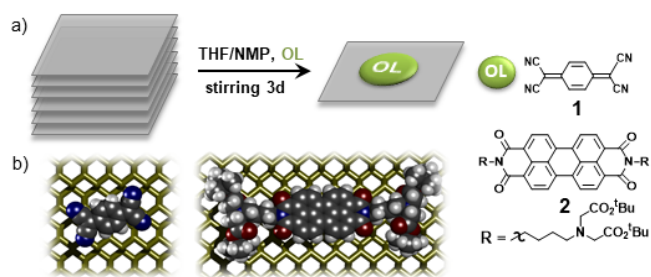

**Figure 4.** a) Representation of the organic ligand (OL) driven exfoliation/functionalization of bulk *BP* with TCNQ (**1**) and perylene bisimide (PDI) (**2**) leading to the formation of charge transfer compounds, consisting of *FL-BP*/*SL-BP* sheets, decorated with strong electron withdrawing *OLs*. b) Schematic representation of (**1**) and (**2**) adsorbed on *BP*.

What the covalent addition chemistry is concerned the situation is different. The 2D sheet polymers are strain free and the driving force for an initial addend binding is actually very low. It is reasonable to assume, although not proven yet, that such initial covalent binding events preferentially or even exclusively take place at pre-existing basal plane defects or at the edges of the sheets, which also can be considered as defects. As a consequence, defect free and non-activated graphene and $MoS_2$ can be considered as rather inert towards addition reaction. However, in BP the situation is different, since phosphorus is highly oxophilic. Indeed, in the case of *SL*-BP chemistry a major challenge is the protection of the exfoliated sheets against oxidative degradation.[25a, 35b, 47]

## 3. The Chemistry of Graphene

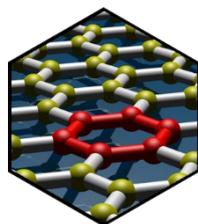

Among the three prototypes of 2D sheet polymers, which are at the focus of this article, the chemistry of graphene is the relatively most advanced. Since a number of recent review articles already appeared,[30] here only briefly the most important accomplishment will be listed.

The unsaturated π-system of graphene forms a rather soft and polarizable surface. This sea of π-electrons does not only favor the energy releasing re-aggregation to graphite but also the non-covalent binding of small molecules. Such favorable interactions are expressed, for example, by the dispersability[10-11, 23c, e] of graphite in "soft" solvents such as *N*-methyl-pyrrolidone (NMP)[10] or in aqueous solutions containing amphiphilic perylenes.[41a, b] These kind of dispersing agents lead to very pronounced exfoliation even down to single layer sheets (graphene). However, it has to be pointed out that the exfoliation is not exhaustive because next to graphene also large fractions of few-layer graphene and multi-layer graphene (graphite) are still present. This is a common drawback for all exfoliated 2D materials, but concepts have been developed to quickly determine the nanosheet size, thickness, and concentration by optical extinction measurements.[48]

In the case of graphite exfoliation with π-surfactants in water, the dispersion process is strongly facilitated by hydrophobic interactions. The non-covalent interactions of graphene with polarizable and π-conjugated organic molecules are conceptually much less pronounced in organic solvents. In this case the re-aggregation to graphite is a process of overwhelming competition. Very well defined assemblies of hybrid architectures can be generated by the sublimation of organic molecules, i.e. perylenes, on surface-supported graphene monolayer.[49] It has also been shown that integrated organic-inorganic hybrid architectures can be build up by a simple layer-by-layer approach.[50] In the first cases the highly ordered hybrid structures were characterized with atomic resolution using STM. The pronounced supramolecular interaction of graphene with the organic dye molecules can be demonstrated, for example, by fluorescence quenching, seen in optical- and Raman-spectroscopy.[41a] This process is based on a pronounced electronic communication of the two components based on photo-induced energy- or electron-transfer processes.[51]

The covalent chemistry of graphene is conceptually based on a) addition reactions of suitable addends to the π-bonds of the basal plane or b) by the binding to the edges. Both scenarios, however, bear inherent difficulties. Firstly, the basal planes are rather unreactive. Secondly, the ratio of edge to basal plane positions is very low. As a consequence, even if an exclusive edge functionalization (e.g. quenching of dangling bonds, substitution of edge functionalities, additions to edge carbon-carbon double bonds) has been accomplished it is very difficult to prove, since no significant spectroscopic changes, for example in the Raman-spectra, can be expected. It is reasonable to assume that basal plane functionalization proceeds preferentially at pre-existing defects.

In general quite reactive addends are required in order to promote a measurable functionalization.[30d] Typical examples are radicals,[30g, 52] diazonium compounds,[30c, e] arynes,[53] carbenes,[54] nitrenes[55] or tetracyano-ethylene[56] which is able to undergo Diels-Alder-reactions. The efficiency of the covalent addend binding can be considerably increased, if the graphene is reductively activated *prior* to the covalent bond formation.[30b] Negatively, charged graphenides and graphite intercalation compounds can easily be generated by treatment with alkali metals.[57] Actually, GICs with alkali metals intercalants represent a well-established compound class that has extensively been studied already in the middle of the last century.[58] The stoichiometry of the intercalation process can easily be monitored by Raman spectroscopy [59] and their structure elucidation has been carried out by angle-resolved photoemission spectroscopy.[60]

The reaction of alkali metal graphenides or GICs with organic electrophiles such as alkyl- or aryliodides,[30g, 52, 61] aryldiazionium compounds[32, 40d, 61d] or aryliodonium compounds[38] can either be carried out in dispersion or as surface supported sheets.[30j, 44b] The reaction of the intermediately formed graphenides with water[42-43] yields polyhydrogenated graphane-like systems with interesting photophysical properties.[39b, 62]

In general compared with the corresponding reaction based on neutral graphene the degree of functionalization can be considerably enhanced in the case of reductively activated starting materials, as demonstrated for example, by Raman-spectroscopy or by TGA-measurements.[36b, 38, 61d] The reaction sequence starts with an electron-transfer from the graphenide, serving as an excellent reducing agent, to the electrophile. After or during the cleavage of either $N_2$ or iodide highly reactive radicals are generated, which are able to attack the carbon atoms of the graphene sheet (Scheme 1).

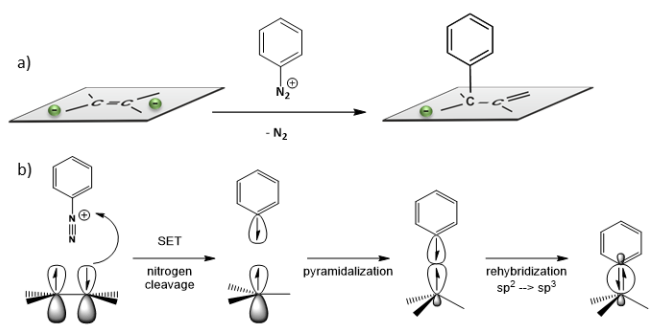

**Scheme 1.** a) Reductive arylation of graphenides – phenyldiazonium cation as electrophile. b) SET mechanism: A single electron transfer leads to the extrusion of $N_2$ and the generation of an aryl radical with subsequent covalent binding to the graphene basal plane.

The most preferred binding process leading to the least strained adducts would be an *antaratopic* 1,2-binding motif (Figure 5).[44a] A *supratopic bis*-addition always leads to rather strained geometries whose formation is unfavorable. This can be demonstrated by the inertness of the outer layer of a bilayer graphene compared with the much more pronounced reactivity of a monolayer on a $SiO_2$ substrate, that is able to *antaratopically* react with graphene dangling bonds formed by an initial addend attack from the opposite side.[44b]

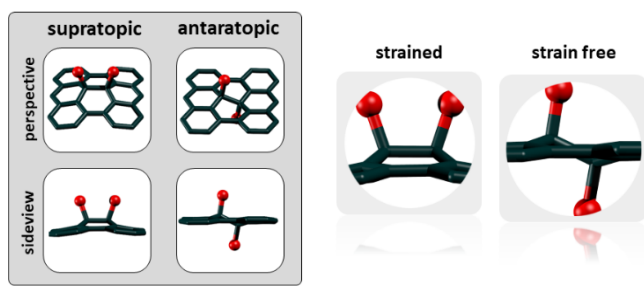

**Figure 5.** *Supratopic* and *antaratopic* 1,2-addition on a sp² carbon plane. In the *supratopic* addition scheme, strain is build up in the respective 2D layer, whereas a strain free situation is achieved in the *antaratopic* case.

## 4. Exploring the Chemistry of Molybdenum Disulfid

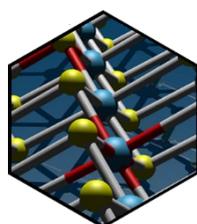

The scientific research on two-dimensional nanosheets of layered transition metal dichalcogenides has unprecedentedly evolved within the last five years and their potential for a of applications i.e. in the field of hydrogen evolution,[64] energy storage,[65] hybrid generation,[29b] and sensing[66] has been elucidated. As outlined in the introduction, MoS$_2$ is one extraordinary example of this novel class of compounds and has received a lot of interest by many researchers working at the borderline between physics, chemistry, and material science. MoS$_2$ can be isolated in two polytypes (Fig. 6) - 2H-MoS$_2$ and 1T-MoS$_2$ - both exhibiting fundamental different physical properties.[21a, 22, 67] Zhang *et al.* have summarized the recent developments in the preparation, the hybridization, and applications of solution-processed MoS$_2$ nanosheets.[23o] In 2013, Dravid and coworkers succeeded in the first chemical modification of chemically exfoliated MoS$_2$ by a ligand conjugation with polyether thiols.[31a] Nevertheless, the covalent modification of this 2D sheet polymer is not a facile and straight forward process as it was repeatedly stated that the basal plane of MoS$_2$ is rather inert.[31b, c]

Conceptual possibilities for covalent MoS$_2$ derivative formation are: a) the formation of addition products (*APs*), where the addends bind to sulfur atoms changing the binding- and/or valence state and b) the formation of substitution products (*SPs*), where sulfur atoms of the MoS$_2$ plane are formally substituted by organothiols (Fig. 7).[31d] The formation of *APs* could be accomplished, for example, by the binding of radicals R· to the plane of sulfur atoms[31b] or by quenching of the negatively charged MoS$_2$ sheets with electrophiles involving a good leaving group.[31c]

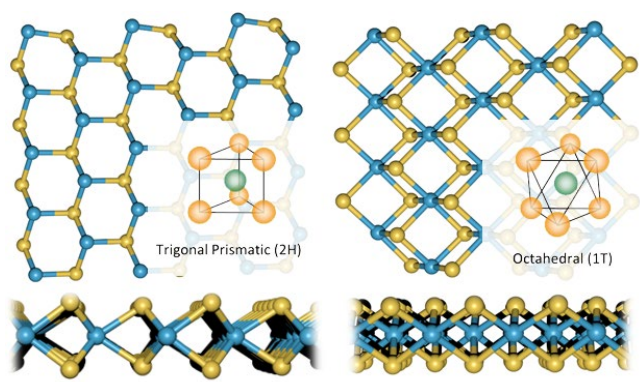

**Figure 6.** Top and side view of the semiconducting 2H-MoS$_2$ (left) and metallic 1T-MoS$_2$ (right) polytypes.

The most promising way to access MoS$_2$ addition products would c) be to react MoS$_2$ sheets containing sulfur vacancies with thiols in order to heal the defects.[31a, e] In all of these cases the binding state of the addend carrying sulfur atoms would be different compared with the parent situation found in MoS$_2$.

First attempts to functionalize MoS$_2$ where carried out with both the 1T- and 2H-polymorphs.[31b, c] A very useful prerequisite for the wet chemical functionalization of MoS$_2$ is a preceding exfoliation of the sheets in a dispersing agent. Here, a successfully applied method is the treatment of bulk MoS$_2$ with *n*-butyllithium which leads to the formation of lithium intercalation compounds,[31a, c, 68] associated with a widening of the MoS$_2$ interlayer distance and a negative charging of the sheets with the intercalated Li$^+$ cations serving as counterions.[68a-c, 69] Pumera and coworkers have tested different organolithium compounds for the exfoliation of MoS$_2$ and screened the catalytic activity of the respective material with respect to a hydrogen evolution reaction.[64b] In general, the lithium intercalation process is also accompanied by the conversion of the trigonal prismatic 2H phase into the octahedral 1T polymorph (Fig. 9a). The resulting material is reasonably stable and can be dispersed in water, which yields the formation of chemically exfoliated MoS$_2$ (*CE*-MoS$_2$). It was stated that the treatment of the exfoliated material with water leads to the development of H$_2$ and the re-oxidation to neutral CE-MoS$_2$.[31a, 68d] Other reports consider the negatively charged *CE*-MoS$_2^{n-}$ sheets to remain in dispersion.[31b, c] For the dispersion of 2H-MoS$_2$, for example, the treatment in viscous solvents such as NMP or CHP[23c-f, 48a] or in isopropanol[70] has been reported. The stability of dispersions in water can be considerably increased if cationic amphiphiles are added.[68d] This allowed for the preparation of MoS$_2$-polymer composites.

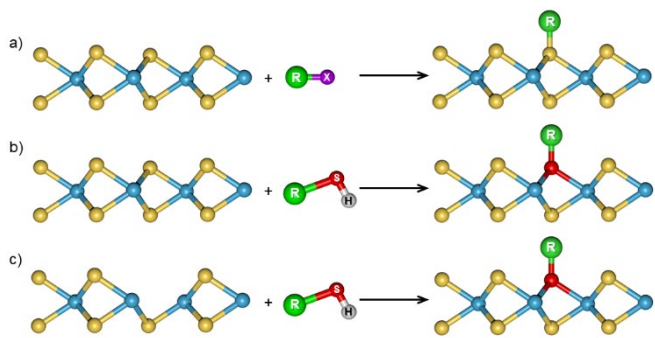

**Figure 7.** Conceptual possibilities for the MoS$_2$ functionalization (2H phase chosen as an example). a) Incoming addends binds to a sulfur atom of the MoS$_2$ layer. b) Formal substitution of a MoS$_2$ sulfur atom by an organic thiol. c) Filling of an existing S vacancy by an organic thiol.

As outlined, it has early been observed that the *n*-butyllithium promoted exfoliation of MoS$_2$, which involves also an ultrasonication step, leads also to the formation of defects (mainly described as sulfur vacancies).[19, 21b, 31a, 69] This on the other hand provided the basis for one of the initial covalent functionalization approaches for MoS$_2$. Here, the treatment of exfoliated MoS$_2$ sheets with thiol-terminated ligands,[31a] lead to a covalent binding of the ligand sulfur atom to the coordinatively unsaturated molybdenum centers and to the surrounding sulfur atoms already present at the defect site (Fig. 8). Theoretical investigations of sulfur vacancy and strain energy containing MoS$_2$ basal planes suggested a catalytic activity of these structures with respect to H-binding and hydrogen evolution reactions (HER).[18b]

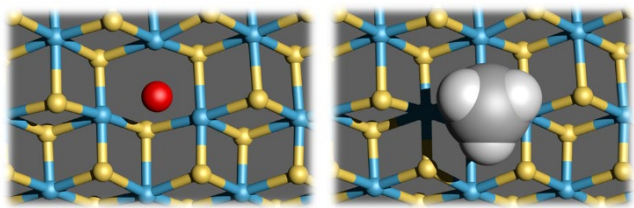

**Figure 8.** Schematic representation of the filling of a sulfur vacancy in the MoS$_2$ lattice by the reaction with thiols – here Me-SH; position of the missing sulfur atom marked in red.

The concept of sulfur vacancy filling was further extended to the binding of a series of commercially available thiols. The morphology of the functionalized 2D sheets was investigated by microscopy (TEM, SEM, AFM), TGA and spectroscopic methods including Raman, FT-IR, and XPS.[71] However, precise atomistic details on the binding mode could not be obtained.

In a related study the electronic influence of both aromatic and aliphatic sulfur bound addends were investigated by photo-luminescence (PL) and photoelectron spectroscopy in air (PESA).[72] A more recent investigation on the vacancy functionalization of MoS$_2$ suggested an alternative reaction sequence. Here, liquid exfoliated 2H-MoS$_2$ was allowed to react with cysteine.[31e] After purification of the reaction product and re-dispersion a modified dispersability in *iso*-propanol was observed indicating nanosheet aggregation. Instead of finding signatures that would support the targeted covalent binding of cysteine, the formation of the oxidized dimer of cysteine, namely cystine, could be shown. Evidence was found that 2H-MoS$_2$ mediates the oxidative formation of cystine and it was demonstrated that air oxygen is not responsible for this process. The same dimerization process occurred also with other commercial thiols. Obviously, the disulfides formed during the exposure to MoS$_2$ are only physisorbed on the surface of the MoS$_2$ sheets. The implications of this careful study are that caution with derivatization scenarios employing organic thiols for the covalent functionalization of MoS$_2$ has to be taken.

On that rationale, the first successful direct covalent functionalization of the sulfur ligand plane has been based on a lithiation approach with an *in situ* generation of chemically exfoliated 1T-MoS$_2^{n-}$ monolayers and a subsequent reaction with 2-iodoacetaminde or iodomethane (Fig. 9).[31b]

During these studies it has been demonstrated that the negative charges on the nanosheet can also be removed by a treatment with I$_2$ without promoting covalent bond formation. The reactions products have been characterized by XPS, IR-spectroscopy, $^{13}$C CP-MAS-NMR-spectroscopy, optical spectroscopy, and TGA. The latter investigations revealed a degree of functionalization of about 20 at%. Similar results were obtained with WS$_2$. After the functionalization the properties of the 1T phase were significantly altered from metallic to semiconducting, giving rise to strong and tunable photoluminescence.

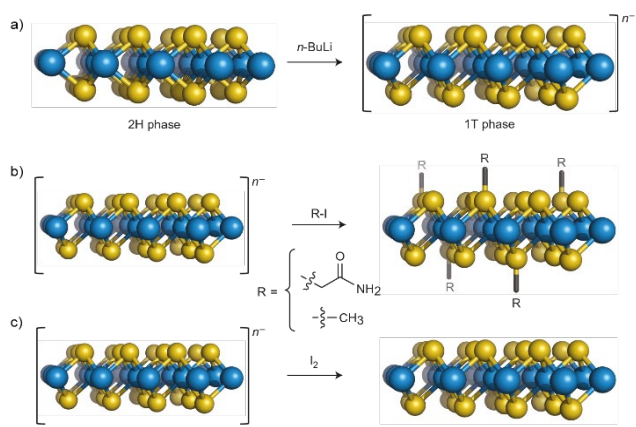

**Figure 9.** a) The 2H phase of TMDs is converted to the 1T phase *via* lithiation using *n*-butyllithium (*n*-BuLi), and the 1T phase is negatively charged; n⁻ indicates the excess charges carried by the exfoliated 1T phase nanosheets. b) The nanosheets are functionalized using 2-iodoacetamide or iodomethane (R-I). c) The charge on the nanosheets can also be quenched by reacting with iodine. Adapted from Ref. [31b] with permission from Nature Publishing Group.

Our own group followed a similar strategy for the covalent functionalization of MoS$_2$ (Fig. 10).[31c] Here, the intermediately generated 1T-MoS$_2^{n-}$ nanosheets have been trapped with diazonium compounds. Typical degrees of functionalization were found to be in the range of 10-20 at% and can be fine tuned by the choice of the intercalation conditions. Annealing at 350 °C restores the pristine 2H-MoS$_2$ phase. The covalently functionalized reaction products are well dispersible in anisole, confirming a significant modification of the surface properties by functionalization. DFT calculations have shown that the grafting of the functional entities to the sulfur atoms of MoS$_2$ is energetically favorable and corroborated the S-C bond formation.

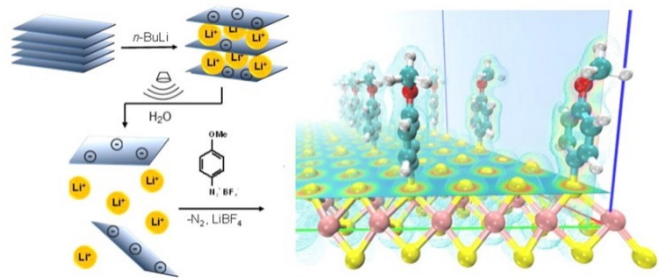

**Figure 10.** Schematic representation of the basal plane functionalization of MoS$_2$. After intercalation with *n*-butyllithium, the negatively charged MoS$_2$ is dispersed in water by mild bath-type sonication leading to an efficient exfoliation into individual sheets. The charges on the MoS$_2$ are quenched by the addition of 4-methoxyphenyldiazonium tetrafluoroborate obtaining the functionalized product. Adapted from Ref. [31c] with permission from American Chemical Society.

Most recently, thiobarbituric acid was used as a nucleophilic thiol reagent to efficiently functionalize the edges and basal planes of chemically exfoliated MoS$_2$ and WS$_2$.[31j] In the MoS$_2$-thiobarbituric acid conjugates, the, metallic character of the respective 1T phase was preserved.

Metallic T-MoS$_2$ can also be functionalized with *para*-substituted iodobenzenes (-OCH$_3$, -H, and -NO$_2$).[73] Here it has been demonstrated that the degree of covalent functionalization significantly depends on the nature of the *para* substituent. The highest degree of functionalization has been determined for the electron withdrawing nitro group. Accompanying first-principles calculations have shown that the benzene derivatives bind much more strongly to the 1T polytype. The work function is considerably tunable with the functionalization of MoS$_2$ and increases the potential for applications in photocatalysis as well as electronic and optoelectronic devices.

Another concept for the chemical functionalization of MoS$_2$ is the treatment of *iso*-propanol exfoliated and dispersed sheets with metal salts (M(OAc)$_2$, M = Ni, Cu, Zn; OAc = acetate).[70] In this case, the reaction of liquid exfoliated 2H-MoS$_2$ nanosheets yields the respective 2H-MoS$_2$-M(OAc)$_2$ derivatives, which have been characterized by XPS, DRIFT-IR, and TGA. The coordinative functionalization leads to an increased solubility in "nonconventional" 2D layered materials solvents like acetone. This concept of coordination chemistry and multidentate ligand binding has first been reported for the functionalization of 3D MoS$_2$ nanoparticles where polymeric chelating ligands were grafted to surface metallated precursors.[74]

## 5. The Advent of the Black Phosphorus Chemistry

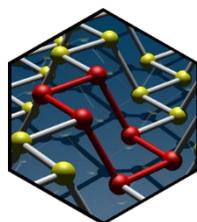

Compared with other inorganic 2D architectures, such as transition metal dichalcogenides, the construction principle of BP is simpler. BP has a non-planar, puckered architecture assembled of fused hexagons of sp$^3$-hybridized P atoms. The aesthetically pleasing lattice structure (Fig. 11) constitutes of a two dimensional σ-only system, involving one lone electron pair at each P atom.

Whereas in recent years many outstanding physical and materials properties of BP have been investigated quite intensively,[16a, 25b, 75] its chemistry remained rather unexplored. Only recently, a first series of non-covalent[31f, 35a, b] and covalent[31g, i] functionalization protocols have been introduced. Why is the exploration of the chemistry of BP so important? We name the following arguments: The exploration of BP chemistry will a) allow to considerably improve the processability and increase the solubility of this nanomaterial; b) it will establish concepts for its chemical stabilization, circumventing its pronounced oxophilicity; c) it will provide the opportunity for very extended modulations and for fine tuning of its physical properties; d) it will give access to the combination of the physical and materials properties of BP with those of other compound classes; e) it will reveal at a fundamental level the intrinsic chemical properties and reactivity principles of BP; and f) it will serve as a basis for the development of practical applications of BP-based materials and hybrid architectures.

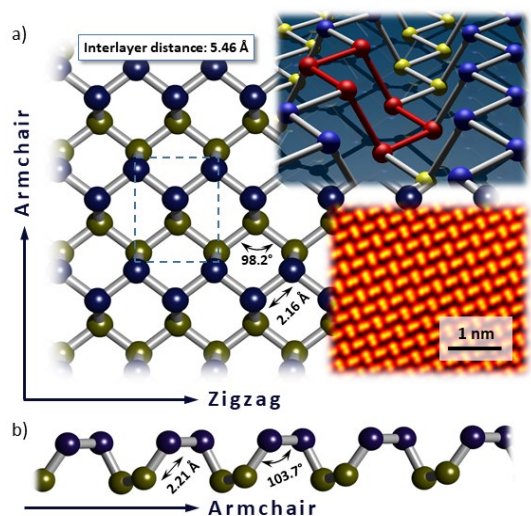

**Figure 11.** a) Top view of the puckered honeycomb lattice of BP – upper plane P atoms marked in blue, lower plane P atoms given in yellow. (b) Lateral view on the lattice in armchair direction. Characteristic distances and angles according to Castellanos-Gomez *et al*.[16b] Insets: BP lattice with six membered ring in chair configuration highlighted in red; Scanning tunneling electron microscopy (STEM) image of the BP lattice.

The most prominent reactivity feature of BP, which however poses challenges towards the exploitation of its chemistry, is the intrinsic instability[35b, 47] of mono- and few-layer nanosheets of BP, which can be obtained by mechanical exfoliation of bulk crystals[16b, 25b] as well as by solvent exfoliation.[13, 23i-n] This instability especially against ambient oxygen and moisture leads to a fast oxidative degradation.[23i, j, m, 47] In addition to that, light irradiation can foster the degradation through photo-oxidation.[47b] Although bulk crystals of BP are more stable, thin flakes of BP below 10-nm thickness degrade in days, whereas single- and few-layer samples may even degrade within hours.[16b, 47a, 75g]

An obvious way to protect BP against oxidative degradation would be the non-covalent coverage of its surface with small electron withdrawing molecules[35a], ionic liquids[35b, 46] or with other inert 2D sheet polymers. Recent approaches in the field of 2D layered material stabilization include $Al_2O_3$-, $TiO_2$-, titanium sulfonate ligand-, polyimide- or aryl diazonium functionalization.[7,19–23] Encapsulation of BP with other 2D materials has been followed by using graphene or hexagonal boron nitride (*h*BN).[19]

With respect to the stabilization of BP, our group has recently also achieved a promising progress.[35a, b] Our studies on the interaction of BP with perylene bisimides (*PDIs*) (**2**) and 7,7,8,8-tetracyano-*p*-quinodimethane (*TCNQ*) (**1**) constituted the first supramolecular chemistry investigations of BP (Fig. 4).[35a] Here, we have demonstrated that the wet chemical treatment of BP with electron deficient aromatic systems like PDIs (**2**) leads to the formation of supramolecular hybrid structures, where the extended conjugated π-system of the perylene undergoes face-to-face stacking with the surface of *FL-BP / SL-BP*. These strong supramolecular interactions lead to the exfoliation of the BP flakes as well as to the formation of a protection layer, all over the surface of the BP, avoiding its oxygen degradation. Even more remarkably, the analogous reaction of BP with the electron acceptor TCNQ (**1**) leads to a pronounced electron transfer from BP to TCNQ generating a two dimensional charge transfer salt consisting of positively charged FL-BP and a negatively charged TCNQ coverage. The positive charge on the BP surface is stabilized by the layers underneath, which is nicely supported by quantum mechanical calculations (controlled oxidation). As a consequence, even the formation of the $TCNQ^{2-}$ dianion is supported, which can easily be monitored by absorption spectroscopy.

For a further very systematic study on the oxidation and passivation of mechanically exfoliated BP,[35b] we have combined AFM, SRM, and SRS as rapid and reliable methodology to precisely estimate the thickness of BP flakes using the Si-attenuation intensity (Fig. 12). Moreover, it has been demonstrated that the intensity ratio of the $A^1_g/A^2_g$ BP-Raman bands are highly indicative for the analysis of the oxidation status. As outlined, the oxidative degradation of BP decreases with increasing flake size and thickness. We were able to show that in the absence of light NMP covered <10 nm thick BP flakes are reaching stabilities of one month by keeping the samples in darkness. Significantly, it could also be demonstrated that the diffusion of oxygen and water as well as the BP photo-oxidation can be suppressed by covering the flakes with the ionic liquid BMIM-BF$_4$ which leads to an outstanding stability of months under environmental conditions.

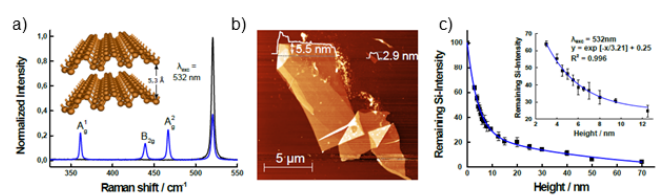

**Figure 12.** a) Representative Raman spectra of BP (blue) and the SiO$_2$/Si substrate. The inset shows the chemical structure of BP, highlighting the interlayer distance of 5.3 Å. b) AFM image of an origami-like BP flake with different terraces and orientations due to its folding. The thinner areas exhibit a thickness of ca. 2.9 nm. c) Normalized silicon intensity attenuation plot measured using different flakes. The inset shows the height below 12 nm. The curve can be adjusted by an exponential decay type I. Adapted from Ref. [35b] with permission from American Chemical Society.

Covalently functionalized SL-BP derivatives were first investigated by theoretical calculations.[25a, c] For this purpose oxy- and imine functionalized SL-BP with either single sided (*supratopic*) or double sided (*antaratopic*) functionalization motifs have been considered.[25a] The DFT calculations suggested that oxy-functionalized BP could be accessible under conditions ranging from ultrahigh vacuum to high concentrations of molecular O$_2$, while the imide-functionalized BP could be formed at relatively high concentrations of hydrazine. A comparative first-principles study[25c] proposed that a single-sided atom-by-atom coverage of SL-BP with monovalent addends such as hydrogen and fluorine results in the formation of regions with sp$^2$-hybridized and non-functionalized P atoms next to the addend carrying sp$^3$-hybridized P atoms. This leads to metallic subdomains. In contrast, a double-sided coverage with H- and F atoms could lead to the formation of metastable 1D addition patterns. Furthermore, it was proposed that a complete hydrogenation and fluorination leads to a decomposition of its structure into narrow PH (PF) chains bonded by hydrogen-like bonds. In general, fully fluorinated BP was found to be more stable than the hydrogenated analog. In contrast to the functionalization with monovalent addends the oxygenation of BP could give rise to a transition into a disordered amorphous 2D system according to these calculations.

The reaction of mechanically exfoliated BP with aryl diazonium compounds led to the first covalent functionalization of this 2D sheet polymer (Fig. 13).[31g] The characterization of the covalently functionalized adducts has been carried out by XPS, Raman spectroscopy, and AFM measurements. The proposed reaction mechanism of the adduct formation consists of a sequence of electron transfer reactions from BP to the diazonium compound, comparable to what was found for graphene.[30c] This leads to an elimination of N$_2$ and the generation of reactive aryl radicals, which subsequently attack the basal BP plane yielding the formation of P-C bonds. The binding state was also investigated by DFT calculations. The covalent attachment of the respective aryl derivatives was found to be thermodynamically favorable but appears to cause a significant lattice distortion. The chemical functionalization leads to an increased stability against oxidative degradation compared with unmodified BP. It alters also the electronic properties of exfoliated BP yielding a strong, tunable *p*-type doping that simultaneously improves the field-effect transistor mobility and on/off ratio.

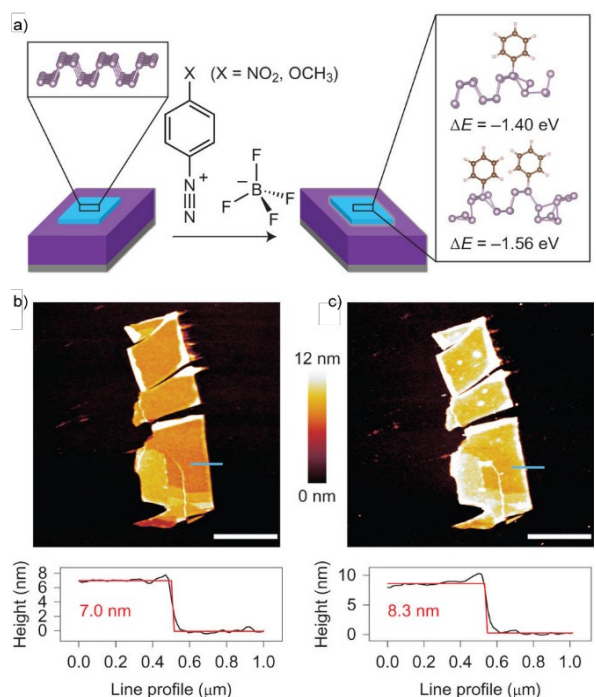

**Figure 13.** a) Reaction scheme of benzene-diazonium tetrafluoroborate derivatives and mechanically exfoliated few-layer BP (light blue) on a Si (grey)/SiO$_2$ (purple) substrate. b) AFM micrograph (top) of a BP flake prior to functionalization, along with the flake-height profile extracted along the blue line (bottom). c) AFM micrograph (top) of the same flake after 30 minutes of exposure to 10 mM 4-NBD. Adapted from Ref. [31g] with permission from American Chemical Society.

With the introduction of the most recent achievements of BP functionalization we would also like to emphasize in the following some major challenges, opportunities, and perspectives in the field of the 2D chemistry of BP.

**5.1. Large Scale Production**

The large scale and large area production of SL-BP remains unsatisfactory and is still in its infancy. Thin flakes of BP (mixtures of SL-BP, BL-BP or FL-BP) were first prepared mostly by mechanical cleavage of bulk crystals.[23i, k, l] This pioneering procedure allowed for the exploration of the fundamental physical properties of BP but is not suitable for a large scale production. Liquid exfoliation using viscous solvents such as *N*-cyclohexyl-2-pyrrolidone (CHP) allows for upscaling and bulk production of FL-BP.[13] Along these lines, we have recently shown, that even rather stable dispersions of flakes with observable photoluminescence can be prepared.[23l] Nevertheless, this approach does not qualify for an easy generation of large area SL-BP, keeping it still an elusive goal. Chemical vapor deposition (CVD) methods that have been very successfully employed in the production of graphene[24a] and other 2D materials like boron nitride[76] would be highly desirable. However, a first attempt in this direction using the conversion of red phosphorus into the thermodynamically more stable BP was reported.[77] Next to establishing protocols for the so far elusive CVD approach possible concepts for large area production of SL-BP could involve the oxidation of surface supported reduced SL-BP$^{n-}$ generated from the *Coulomb* force driven exfoliation and deposition of AM-BPICs (AM = alkali metal, BPIC = BP intercalation compound). As will be pointed out below the investigation of AM-BIPCs is also of great interest for a series of further aspects. This concept has precedent in the successful exfoliation of graphite intercalation compounds (*GICs*) into SL-graphenides with subsequent reoxidation, yielding graphene.[42, 57, 63]

## 5.2. Black Phosphorus Intercalation Compounds (BPICs)

In contrast to graphite where the intercalation of guest systems, such as alkali metals or small inorganic molecules between the graphene sheets yields a large variety of intercalation compounds, that have been investigated in great detail for many years,[58b, 59-60] the field of BP intercalation compounds (BPICs)[78] – Fig. 14 - remained almost completely unexplored for a long time.[79]

It is clear that the interfacial spacing between two individual BP layers with about 5.23 Å is roughly 2 Å larger than that in graphite and this provides an ideal prerequisite for the uptake of small guest systems. Theoretical investigations[80] on BPICs have shown exciting opportunities for developing high capacity sodium or lithium ion batteries with up to ~2,600 mAh g$^{-1}$ outperforming graphite based LIBs. For the intercalation of lithium (Li) into FL-BP superconductivity with $T_c$ of 16.5 K has been predicted.[81] Two families of BPICs are especially attractive namely, i) AM-BPICs (AM = alkali metal) and ii) BPICs containing small molecules. Theoretical calculations have predicted that in Na-BPICs the layered structure of BP remains intact up to a maximum loading stoichiometry of 0.25 equiv. sodium (Na) relative to P.[82]

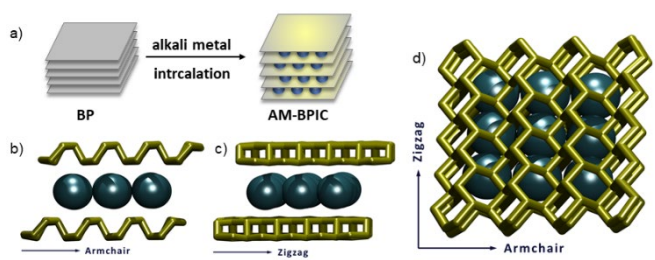

**Figure 14.** a) Graphical representation of the intercalation of BP by alkali metals: formation of AM-BPICs. b) Side view in armchair direction. c) Side view in *zigzag* direction. d) Top view on the alkali metal intercalated BP.

Very recently, some experimental evidence on BPICs was obtained by *in situ* TEM.[83] Also, the intercalation of BP with several alkali metals (Li, K, Rb, and Cs) and alkali-earth Ca was reported all of which exhibited a universal superconductivity with a critical temperature of *ca*. 3.8 K.[79] However, the intercalation phases were mainly superficial (*ca*. 10 microns) and no detailed structural or spectroscopic information was provided. A subsequent study has investigated in detail the vapour phase synthesis and characterization of bulk BPICs with alkali metals (K and Na).[35c] Here, the intercalation process was monitored by *in situ* XRD The obtained results were complemented by DFT calculations which corroborated that the stacking order in bulk BP changes from AB to AC (rather than to AA, in contrast to GICs). This is reflected in the formation of new intercalated phases accompanied by a gliding of the BP layers. The experimental and computational data revealed the maximum layer-structure keeping stoichiometry (≥ 1:4 M:P), from which the layers start to break into chains. Moreover, by a vapour transport intercalation process under ultra-high vacuum followed by *in situ* Raman spectroscopy, a series of novel Raman modes ascribed to the formed BP intercalation compounds have been discovered.

What the small molecule containing BPICs is concerned, so far, the only literature precedent for this kind of fundamental hybrid architecture is the system BP-$I_2$ that has been published almost thirty years ago in a tentative form.[84] It would be of great interest, however, to create and explore many more hybrid architectures including those with inert molecules such as noble gases (He, Ne, Kr, Xe) and $CH_4$ as interlayer guests. Whereas in these cases mostly the electronic decoupling of the basal planes are in the foreground the usage of $H_2$ and $CO_2$ would allow for investigating technologically relevant storage and/or conversion concepts. Also the intercalation of more reactive molecules such as $SO_2$, $SO_3$, $NH_3$, and $H_2O$ could be considered.

## 5.3. Considerations Towards Covalent Functionalization of Black Phosphorus

As the most suitable potential concepts for the covalent chemistry BP the functionalization of neutral BP and reduced BP (BP-ite) have to be considered. In each case both bulk BP and BPICs as well as the corresponding monolayers (SL-BP and SL-BP-ite) supported on surfaces could serve as starting materials. In the cases of bulk functionalization the exfoliation down to FL-BP and SL-BP sheets will be facilitated during the addend binding, causing reduced inter-sheet attractions. In the case of BPIC functionalization this effect will be further enhanced by an electrostatic repulsion of the negatively charged BP-ite sheets. Possible covalent products obtained from the addition of electrophiles such as carbenes, nitrenes, "O" and Lewis acids to neutral BP are depicted in Fig.15. In the first case the direct formation of phosphonium ylid- (**3**) and iminophosphorane (**4**) functionalities within the BP lattice is expected. This may be accomplished by the treatment of BP with carbenes and nitrenes formed *in situ*. The phosphonium ylids (**3**) formed after carbene addition are closely related to *Wittig* substrates.

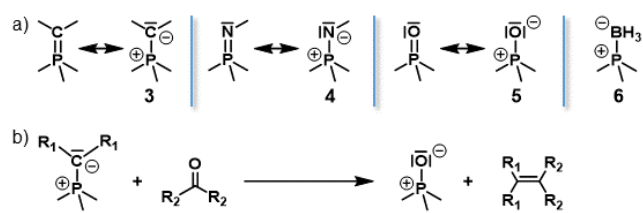

**Figure 15.** a) Covalent addition products: Phosphonium ylid (**3**), iminophosphorane (**4**), phosphinoxide (**5**), and *Lewis* acid adduct (**6**). b) *Wittig* type reaction of BL-phosphonium ylid (**3**).

The significant difference is that three P ligands instead of three phenyl groups are attached to the phosphonium center. This should lead to an increased stability since the positive formal charge on P can be stabilized by the lone pairs of the adjacent P atoms. The reaction of BP with *Lewis* acids is expected to lead to phosphonium functionalities (**5**) within the P lattice. The negative counter charges are located at the addend. The highly bipolar and zwitterionic binding motif is expected to increase the solubility and processability of such BP derivatives considerably.

The functionalization of SL-BP supported on a surface (Fig. 16) would offer the advantage that a) no exfoliation step during the chemical functionalization is required and b) that attacks of the addends can take place only from one side of the phosphorus lattice (*supratopic* addition). Consequently, a very attractive scenario for systematic model investigations is provided, whereas the bulk chemistry will be key for the production of BP derivatives in macroscopic quantities.

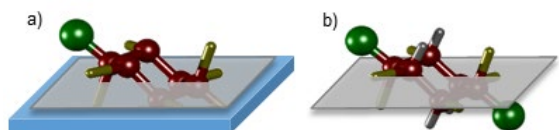

**Figure 16.** a) SL-BP supported on a surface: only a *supratopic* addition is possible. b) SL-BP in solution: *supra-* and *antaratopic* addition reactions are possible.

As with graphenides,[57] compared with neutral BP the reactivity of BPICs towards electrophiles (*E*) is expected to be considerably increased (Fig. 17), since the energetically high lying conduction band is now occupied with electrons. Like in the established chemistry of graphite intercalation compounds[32, 42-43] the first step of a functionalization sequence is expected to be a single electron transfer from the negatively charged BPIC to *E*. Subsequently, a leaving group such as $N_2$, iodide or bromide can be eliminated from the electrophile leaving behind a radical R• that can then attack the black phosphorite (BP-ite) sheet containing one electron less than the starting material. The success of this diazonium functionalization approach has recently been demonstrated for neutral BP.[31g]

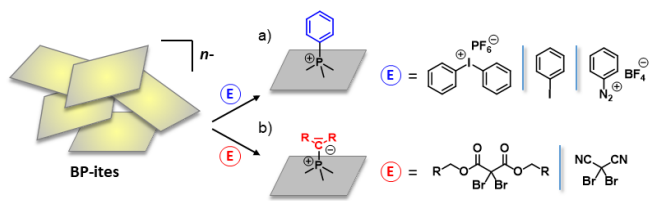

**Figure 17.** Reductive functionalization of black phosphorites (BP-ites). a) Electrophilic trapping with aryl iodonium / diazonium salts and aryl/alkyl-halides – generation of SL-BP phosphonium salts. b) Electrophilic trapping with geminal dihalogenides – generation of SL-BP phosphonium ylids.

If *E* contains two leaving groups the process can take place twice. The covalent functionalization of SL-BP-ites could be accomplished after supporting SL-BP on a substrate and subsequent treatment with an alkali metal. Subsequently, the same chemical transformations as those described for the bulk functionalization could be targeted. Since no accompanying exfoliation process is required, systematic studies on the nature and the regiochemistry and regioselectivity of the covalent functionalization reactions are possible. Moreover, in contrast to the liquid based derivatization scenario only a *supratopic* and no *antaratopic* addition motifs can be realized in the latter case.

### 5.4. General Reactivity Considerations

As already mentioned, the activated BPICs and SL-BP-ites are expected to be more reactive than the respective neutral forms. Since the amount of negative charging of the BPICs and SL-BP-ites can be controlled by the relative amount of the reducing agent (e.g. amount of alkali metal), the reactivity of the starting material could be varied in a rather large and continuous range. The reactivity of BP as a function of the negative charging level, the employed electrophiles *E* as a function of the corresponding redox-potential, and the *Lewis* acid as a function of *Lewis* acidity will be reflected by the obtained degree of addition of the final product, which could be determined, for example, by Raman spectroscopy and TGA/GC/MS.[38]

### 5.5. General Topological Considerations

A fundamental question with respect to the covalent BP chemistry is the regio-regularity of the subsequent additions. In BP the P atoms are sp$^3$-hybridized and trivalent. The addend carrying P atoms in covalently functionalized BP derivatives are also sp$^3$-hybridized but tetravalent. The tetra-valency is true for all binding situations of the derivatives depicted in Fig. 15a. This means, that not only the phosphonium sites but also the phosphonium ylide and phosphinoxide sites involve tetravalent P atoms. The P-C and P-O bonds have no double bond character, but involve *Coulomb* binding contributions due the polarization of the bond (positive formal charges on P and negative formal charges on C and O). In contrast to conjugated π-systems the P atoms in BP are in a first approximation electronically decoupled. Of course, inductive effects of addend carrying P binding sites can have directing influences on subsequent additions. The major driving force for the selectivity of subsequent addition reactions is expected to be due to steric repulsions of the addends themselves and the most even strain distribution of the σ-system within the two-dimensional P lattice. In Fig. 18 series of characteristic spatial binding patterns of subsequent addend binding relations are outlined.

What the binding distance between is concerned we define *syn-g$_{0,0}$* for the binding within the same six-membered P ring and *syn$_{n,m}$* or *syn-g$_{n,m}$* for the binding within different P rings (*g* denotes *gauche*, *n* denotes the row-position of the second addend carrying six-membered P ring and *m* the column position of the second addend carrying P ring, see Fig. 18a. In analogy, we define the corresponding *anti*-binding modes. If the spatial relationship between two addends is *syn-g$_{0,0}$* sterical repulsion can occur. If the binding is non-linear within one row of six-membered P rings such as $syn\text{-}g_{n,0}$ or ($n > 0$) or $anti\text{-}g_{n,0}$ ($n \geq 0$) unsymmetrical and energetically unfavorable distortions of the P lattice are expected. What the most favorable strain distribution of the P lattice is concerned, subsequent $syn\text{-}_{n,0}$– or $anti\text{-}g_{n,0}$ are expected to be preferred. The same considerations hold true for the second dimension by regularly and symmetrically functionalizing the neighboring columns of P rings (Fig. 18b). The structure of addition patterns could be determined, for example, by scanning-probe microscopy.

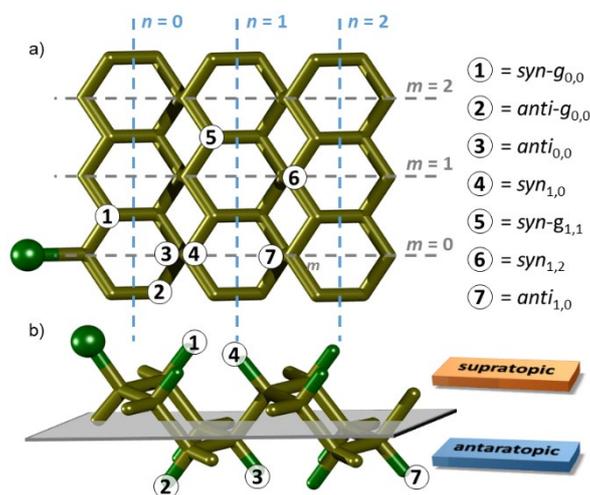

**Figure 18.** BP/covalent addend binding – topological considerations: a) Top view of the BP lattice with one covalently bound addend – spatial binding patterns for a second addend. Relation and position (row/column) are denoted with *syn*/*anti* and indices $n$ (column), $m$ (row). b) Side view of the BP lattice bearing one addend. Second addition sites in the same row are depicted.

If the relative binding topology of two addends has taken place from the same side of the P lattice plane (*supratopic*) their spatial relationship is defined as *syn*. For the binding from opposite sites (*antaratopic*) their relationship can be defined as *anti*.

## 6. Conclusion and Outlook

The chemistry of 2D materials and in particular that of single layer 2D sheet polymers is currently an emerging field at the interface of synthetic chemistry, physics, and materials science. Both covalent and non-covalent functionalization of sheet architectures such as single layer graphene, $MoS_2$, and BP conceptually allows for systematic properties modification such as an improvement of solubility and processability, prevention of re-aggregation, band gap tuning, and combination of properties with those of other compound classes. Next to the examples of 2D sheet polymers mentioned above and serving as role models within this review many other compound classes, e.g. BN, transition metal dichalcogenides in general, are already known. Also the exploitation of their 2D functionalization remains an exciting task of future materials chemistry. However, a number of fundamental challenges have to be addressed. These include the insolubility and polydispersity of most 2D sheet polymers, the development of suitable characterization tools, the identification of effective binding strategies, the chemical activation of the usually rather unreactive basal planes for covalent addend binding, and regioselective plane functionalization. Although a number of these questions remain elusive in this review the first promising concepts to overcome such hurdles have been listed. For all model cases described here, namely graphene, $MoS_2$, and BP, the known schemes for both non-covalent and covalent functionalization have been described. Whereas the covalent and non-covalent chemistry of graphene has already been explored in quite some detail the functionalization of other 2D materials has just been starting to emerge. Among the most powerful characterization tools for functionalized 2D sheet polymers statistical Raman spectroscopy, TGA/MS, optical spectroscopy, scanning probe microscopy, and high resolution electron microscopy have been identified. Already new 2D sheet polymers have appeared, still awaiting their chemical functionalization. These include antimony[17, 85] and germanane.[15, 86] The latter represents the stable and easily accessible homologue of the still unknown graphane, which would be the completely hydrogenated graphene.

**Acknowledgements**

The authors thank the Deutsche Forschungsgemeinschaft (DFG-SFB 953 "Synthetic Carbon Allotropes", Projects A1) and the European Research Council (ERC Advanced Grant 742145 B-PhosphoChem) for financial support. The research leading to these results was partially funded by the European Union Seventh Framework Programme under grant agreement No. 604391 Graphene Flagship.